\documentclass[journal=pcrhej,manuscript=letter]{achemso}
\usepackage[T1]{fontenc}
\usepackage{chemformula}
\usepackage[version=3]{mhchem}
\usepackage[compact]{titlesec}
\usepackage[squaren]{SIunits}
\usepackage{threeparttable}
\usepackage{balance}
\usepackage{color}
\usepackage{lipsum}
\usepackage{achemso}
\setkeys{acs}{maxauthors = 0}
\usepackage{CJKutf8}
\usepackage{lineno}
\usepackage[squaren]{SIunits}
\usepackage{amsmath}
\usepackage{caption,subcaption}
\usepackage{setspace}
\usepackage{multirow}
\usepackage{threeparttable}
\usepackage{booktabs}
\usepackage{graphicx}
\usepackage{dcolumn}
\usepackage{bm}
\usepackage[T1]{fontenc}

\definecolor{americanrose}{rgb}{1.0, 0.01, 0.24}
\definecolor{coralpink}{rgb}{0.97, 0.51, 0.47}
\definecolor{ao(english)}{rgb}{0.0, 0.5, 0.0}
\definecolor{darkpastelgreen}{rgb}{0.01, 0.75, 0.24}
\definecolor{cyan(process)}{rgb}{0.0, 0.72, 0.92}

\author{Yang Xiao}
\affiliation{MIIT Key Laboratory of Semiconductor Microstructure and Quantum Sensing, Department of Applied Physics, School of Physics, Nanjing University of Science and Technology, 210094 Nanjing, China}
\altaffiliation{Contributed equally to this work}
 
\author{Jun-Rong Zhang}
\affiliation{MIIT Key Laboratory of Semiconductor Microstructure and Quantum Sensing, Department of Applied Physics, School of Physics, Nanjing University of Science and Technology, 210094 Nanjing, China}
\altaffiliation{Contributed equally to this work}
 
\author{Sheng-Yu Wang}
 \affiliation{MIIT Key Laboratory of Semiconductor Microstructure and Quantum Sensing, Department of Applied Physics, School of Physics, Nanjing University of Science and Technology, 210094 Nanjing, China}

\author{Weijie Hua}
\email{wjhua@njust.edu.cn}
\affiliation{MIIT Key Laboratory of Semiconductor Microstructure and Quantum Sensing, Department of Applied Physics, School of Physics, Nanjing University of Science and Technology, 210094 Nanjing, China}

 \title{
 Global Impact and Balancing Act: Deciphering the Effect of Fluorination on B1s Binding Energies in Fluorinated $h$-BN Nanosheets
}
\begin{document}

\begin{figure*}[!htb]
\centering
 \includegraphics[width=12cm]{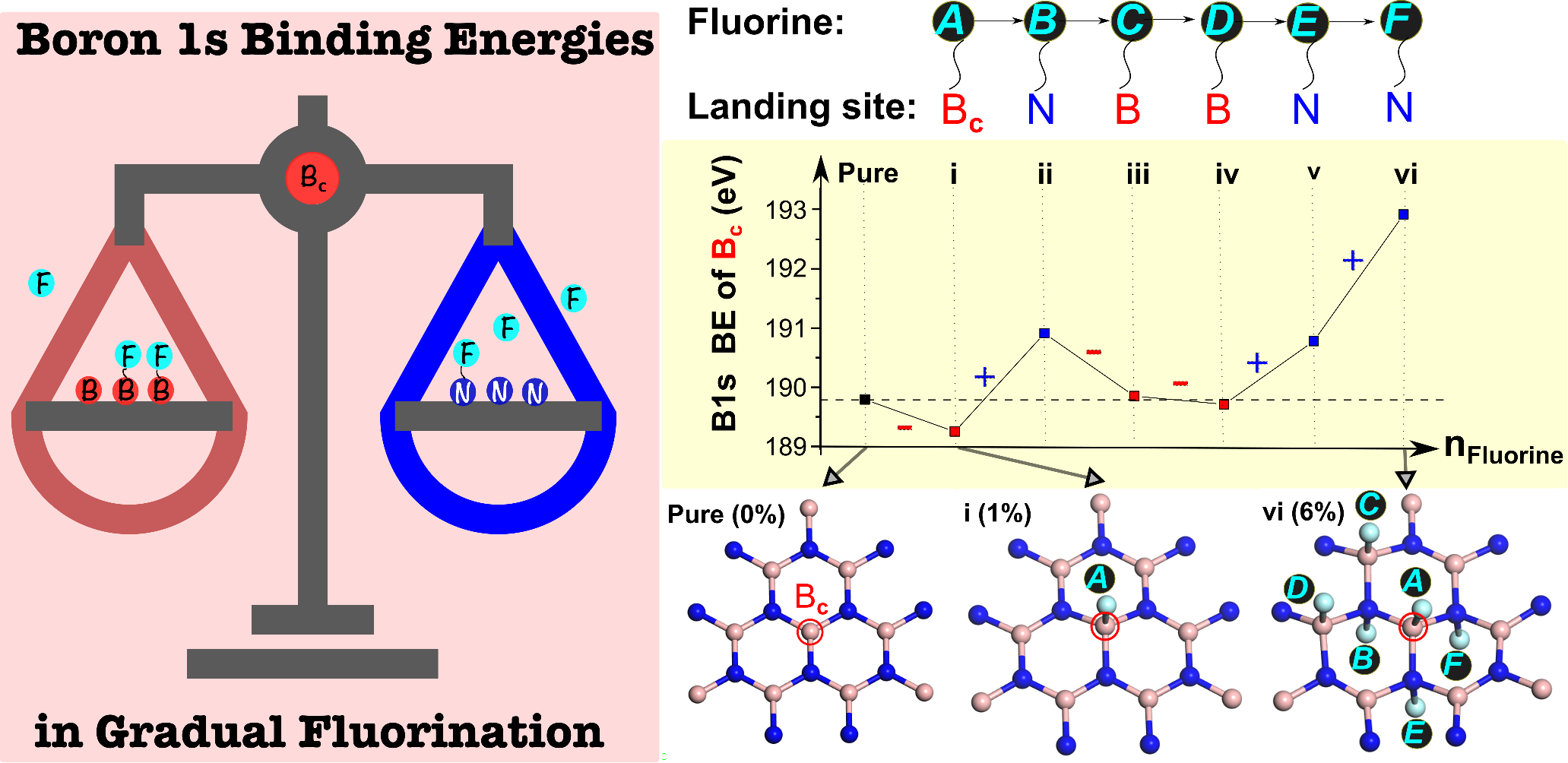}  
 \caption*{\textbf{TOC Graphics}
 }
\end{figure*}

\clearpage
\begin{abstract}
X-ray photoelectron spectroscopy (XPS) is an important characterization tool in the pursuit of controllable fluorination of two-dimensional hexagonal boron nitride ($h$-BN). However, there is a lack of clear spectral interpretation and seemingly conflicting measurements exist.    To discern the structure-spectroscopy relation, we performed a comprehensive first-principles study on the boron 1s edge XPS of fluorinated $h$-BN (F-BN) nanosheets. By gradually introducing 1--6 fluorine atoms into different boron or nitrogen sites, we created various F-BN structures with doping ratios ranging from 1-6\%.  Our calculations reveal that fluorines landed at boron or nitrogen sites exert competitive effects on the B1s binding energies (BEs), leading to red or blue shifts in different measurements. Our calculations affirmed the hypothesis that fluorination affects 1s BEs of all borons in the $\pi$-conjugated system, opposing the transferability from $h$-BN to F-BN.  Additionally, we observe that BE generally increases with higher fluorine concentration when both borons and nitrogens are non-exclusively fluorinated.  These findings provide critical insights into how fluorination affects boron's 1s BEs, contributing to a better understanding of fluorination functionalization processes in $h$-BN and its potential applications in materials science.
\end{abstract}

\clearpage
Fluorine is the element with the highest electronegativity in the periodic table. Fluorination can dramatically alter the properties of diverse materials, such as boron nitride nanosheets\cite{radhakrishnan2017fluorinated, ul_ahmad_novel_2019, venkateswarlu2019effective, zhou_electronic_2010, ahmad_experimental_2020, zhang_first-principles_2019, du_one-step_2014}, boron nitride nanotubes\cite{tang_fluorination_2005, li_fluorination-induced_2008}, and carbon-based materials\cite{touhara_property_2000, bischof_chemical_2023, nanse_fluorination_1997, barbosa_fluorination_2014}, thereby enabling the acquisition of novel chemical and physical attributes.  Fluorination has been demonstrated to be an effective strategy to transition hexagonal boron nitride ($h$-BN), also often referred as ``white graphene'', into a wide-bandgap semiconductor showcasing unconventional magnetic properties\cite{radhakrishnan2017fluorinated, ul_ahmad_novel_2019, zhou_electronic_2010, ahmad_experimental_2020, zhang_first-principles_2019, du_one-step_2014}. The flexibility of this modulation lies both in the degree of fluorination and the fluorination sites, which can occur at both B and N atoms\cite{radhakrishnan2017fluorinated}. Structurally, fluorination at either site changes its bonding type from sp$^2$ to sp$^3$ hybridization. Achieving controllable fluorination is a subject of keen interest to material scientists.

Given the structural intricacy of fluorinated $h$-BN (F-BN), there is a requirement for robust and sensitive techniques for effective characterization. One commonly employed method for the characterization of fluorination in 2D materials is through X-ray photoelectron spectra (XPS) at the B\cite{ul_ahmad_novel_2019, ahmad_experimental_2020, radhakrishnan2017fluorinated, du_one-step_2014, weng_functionalized_2016, venkateswarlu2019effective}, N\cite{ul_ahmad_novel_2019, ahmad_experimental_2020, radhakrishnan2017fluorinated, du_one-step_2014, weng_functionalized_2016, venkateswarlu2019effective}, and F\cite{ul_ahmad_novel_2019, ahmad_experimental_2020, du_one-step_2014, venkateswarlu2019effective} K-edges. This technique, grounded on core excitations, offers element-specific insights and sensitivity towards varying local bonding environments, which are represented by core binding energy (BE) values. For instance, XPS was instrumental in characterizing the dominant nitrogen doping sites in 2D graphdiyne nanosheets.\cite{zhao_few-layer_2018, Ma_carbon2019} 

\begin{figure}[htpb]
\centering
 \includegraphics[width=8.5cm]{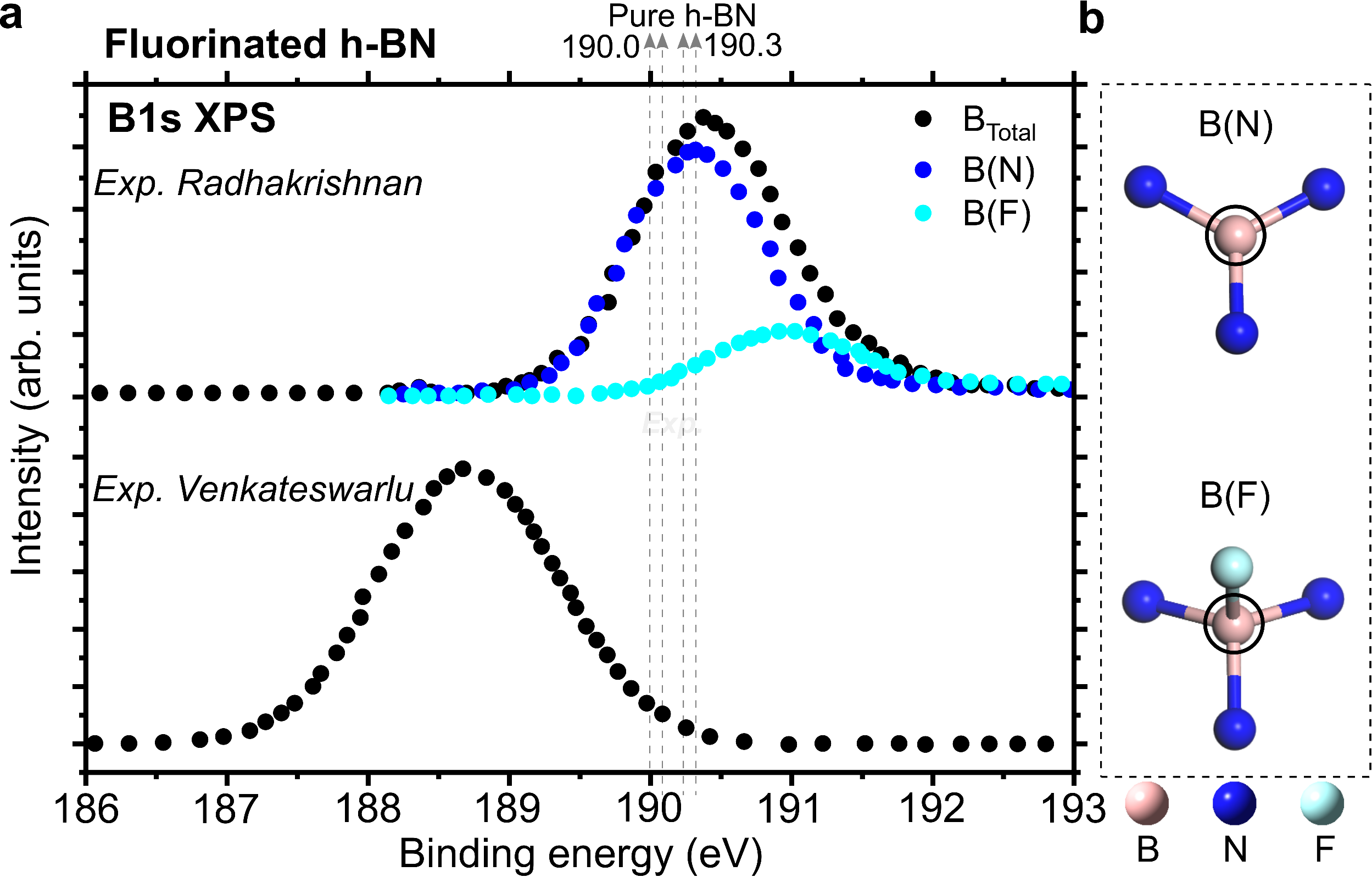}  
 \caption{(a) Recapture of two representative B1s XPS experiments of F-BN (black dots)  from Radhakrishnan et al.\cite{radhakrishnan2017fluorinated} and Venkateswarlu et al.\cite{venkateswarlu2019effective}. \textit{Top}, deconvoluted curves, interpreted as from unfluorinated [B(N); blue dots] and fluorinated [B(F); cyan dots] borons.\cite{radhakrishnan2017fluorinated} Vertical dashes indicate B1s BEs of pure $h$-BN from recent experiments (190.0,\cite{guerra_2d_2018,achour_plasma_2017} 
 190.1,\cite{zhou_high_2014, huang_exponentially_2021} 
 190.22,\cite{deshmukh_ultrasound-assisted_2019} 
 and 190.3 eV.\cite{yu_thermal_2016, yang_thionyl_2021}). 
 (b) Local structures of the two types of borons (in circles).}
\label{fig:exp}
\end{figure}

Nonetheless, the spectral interpretation for F-BN remains unclear due to the absence of direct XPS simulations. This uncertainty extends to understanding the influence of fluorination on $h$-BN \cite{radhakrishnan2017fluorinated, ahmad_experimental_2020, ul_ahmad_novel_2019, venkateswarlu2019effective}. Figure \ref{fig:exp} revisits the two representative kinds of  B1s XPS spectra of F-BN from recent experiments\cite{radhakrishnan2017fluorinated, venkateswarlu2019effective}, as referenced to various pure $h$-BN measurements (190.0--190.3 eV)\cite{guerra_2d_2018, achour_plasma_2017, achour_plasma_2017, huang_exponentially_2021, deshmukh_ultrasound-assisted_2019, yu_thermal_2016, yang_thionyl_2021}. From the broad total peak (top, black dots) at 190.37 eV, Radhakrishnan et al.\cite{radhakrishnan2017fluorinated} resolved two peaks at 191.0 eV (cyan) and 190.3 eV (blue), interpreted as from fluorinated (sp$^3$) and unfluorinated (sp$^2$)  borons [denoted as B(F) and B(N)], respectively. Similar spectra and interpretation were supported by, for example, Ahmad et al.\cite{ahmad_experimental_2020, ul_ahmad_novel_2019}. However, Venkateswarlu et al.\cite{venkateswarlu2019effective} detected a much different total spectrum centered at 188.68 eV (bottom, black dots), which was shifted by ca. --1.69 eV to Radhakrishnan et al.\cite{radhakrishnan2017fluorinated} The total spectrum was assigned to be purely B(F) contributions. The two representative observations show conflicts. As we know, existing XPS data suffer from inconsistency due to different calibration procedures used.\cite{greczynski_x-ray_2020, zhang_choice_2022} Is this conflict a simple calibration problem or is there any deeper physical/chemical insight behind it? Further, in comparison to pure $h$-BN, why does fluorination sometimes lead to a blue (as in Ref. \citenum{radhakrishnan2017fluorinated}) and sometimes a red (as in Ref. \citenum{venkateswarlu2019effective}) shift?  It is crucial to identify the corresponding structural changes associated with these spectral shifts.

Another significant observation that merits discussion is related to distant borons from the fluorination center. Very recently, Bischof et al.\cite{bischof_chemical_2023} proposed that fluorination impacts \emph{all} atoms within this $\pi$-conjugated sheet, meaning, even B atoms distanced from the fluorination center experience the effects of fluorination since this influence is propagated through the system's $\pi$ bond. Hence, a noticeable 1s BE difference is anticipated for an unfluorinated boron in both F-BN and $h$-BN structures. While it is generally recognized that core excitations are localized, the extent of localization varies depending on the system. It was verified\cite{zhang_choice_2022} that the $\pi$ electrons in pure $h$-BN are more localized as compared to graphene or C$_3$N. It is interesting to know whether fluorination may enhance the delocalization and impact distant borons. Thus, it underscores the necessity for precise calculations to aid in XPS interpretations.

Historically, theoretical XPS studies on materials mainly focused on relative BEs.\cite{bagus_consequences_2016, ge2022qm} In recent years, new theoretical methods\cite{ozaki_absolute_2017, zhang_accurate_2019, kahk2021core, brigiano_peculiar_2022} have been developed to calculate absolute BEs, based on large clusters and/or periodic cells. One such method is the cluster-periodic method (CPM),\cite{zhang_accurate_2019, zhang_choice_2022}  which predicts absolute BEs referenced to the Fermi level and allows for direct comparisons with experimental results. CPM is grounded on a full core hole (FCH) density functional theory (DFT) calculation of a large cluster in the gas phase to determine the ionic potential (IP), $I^\text{gas}$, and a ground-state (GS) work function (denoted by $\phi$) calculation based on the periodic cell.  This method bypasses the issue of spurious charges\cite{ozaki_absolute_2017, vines_review_2018} in periodic calculations, triggered by replicas of core holes across different cells. Mathematically, core IP of the solid 2D material is given by\cite{zhang_accurate_2019},
 \begin{equation}\label{eq:cpm}
I^\text{solid} = I^\text{gas}  - \phi.
\end{equation}
Accuracy of the CPM method has been validated through successful XPS spectral simulations of various 2D materials like g-C$_{3}$N$_{4}$\cite{zhang_accurate_2019}, graphene\cite{zhang_choice_2022}, graphynes\cite{zhang_structural_2023, zhang_identification_2023}, and graphdiynes.\cite{Ma_carbon2019}   All early applications were limited to fixed geometries, either using pure 2D materials\cite{zhang_accurate_2019, zhang_choice_2022} or incorporating simple point defects\cite{Ma_carbon2019, zhang_structural_2023, zhang_identification_2023}. Notably, it has been observed that the CPM method is effective in analyzing the electronic structure of pure h-BN, due to the pronounced locality of $\pi$-electrons and the insensitivity to the model shape\cite{zhang_choice_2022}. Thus, $h$-BN serves as a robust template for reliable analysis of more complex substituted structures.

In this Letter, we aim to investigate the influence of fluorination on $h$-BN at different sites (B or N) and varying concentrations (1-6\%) to the B1s BE.  The goal is to gain a comprehensive understanding of the relationship between fluorination and XPS binding energy,  providing valuable insights for understanding fluorination in $h$-BN  and resolving existing debates.

\begin{figure*}[!htb]
\centering
 \includegraphics[width=16.5cm]{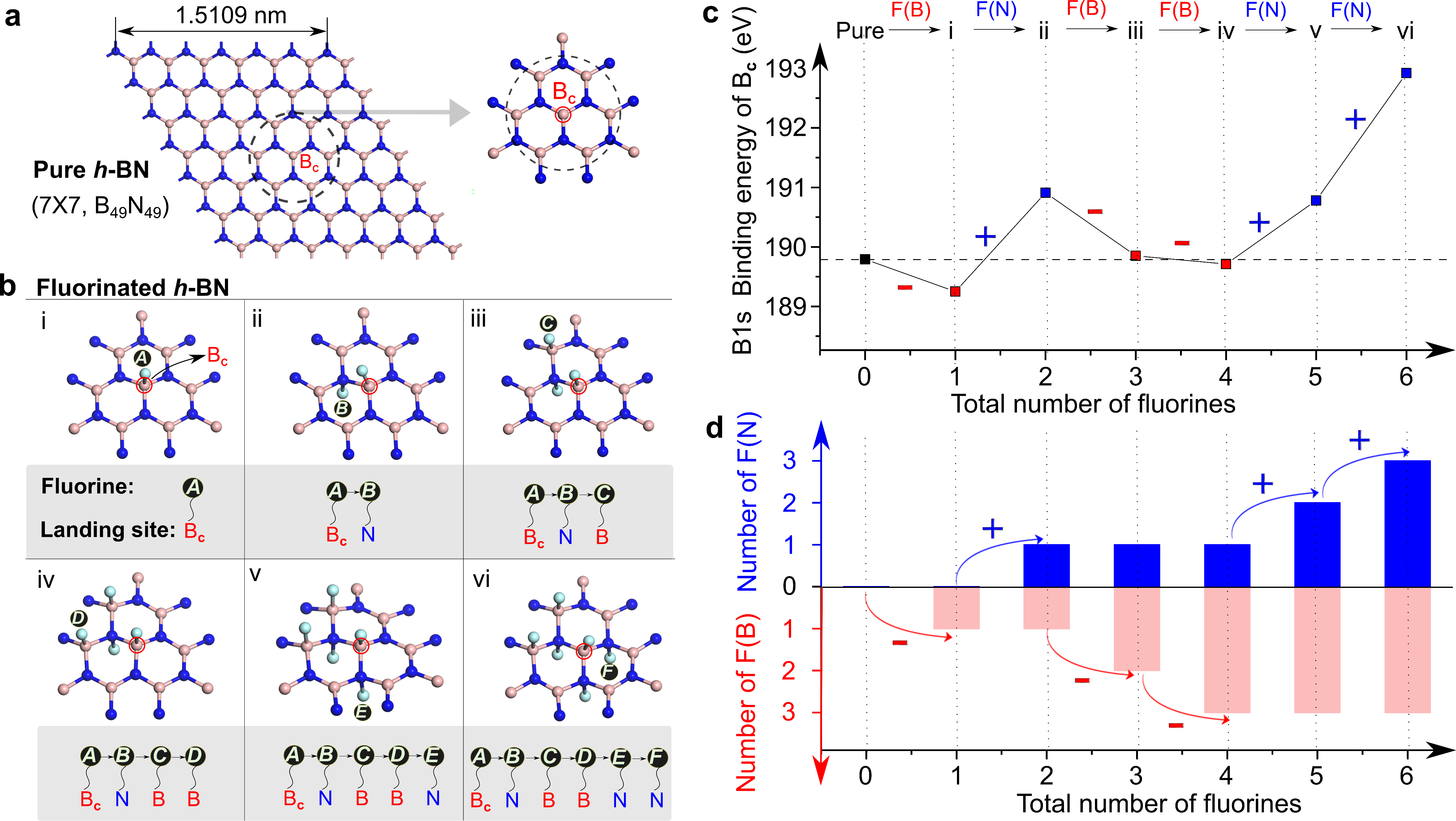}  
 \caption{
 Simulated B1s BEs of B$_\text{c}$ (the central boron) upon gradual fluorination.  (a) 7$\times$7 supercell of pure $h$-BN with enlarged view near B$_\text{c}$.
 (b) Constructed F-BN configurations i-vi (only the central part is shown) by gradual fluorination\cite{radhakrishnan2017fluorinated} with 1-6 F atoms ($A$-$F$), starting from B$_\text{c}$. (c) Simulated B1s BEs at B$_\text{c}$  of i-vi.  The horizontal dash denotes the theoretical BE of pure $h$-BN (189.84 eV).   (d) Number of F atoms landed on nitrogens and borons [denoted by F(N) and F(B), respectively] upon gradual fluorination.  Arrow in blue/pink indicates the positive/negative change caused by the newly landed F(N)/F(B) atom.}
\label{fig:Bc:fluo}
\end{figure*}

The CPM method\cite{zhang_accurate_2019, zhang_choice_2022} was employed to do all B1s BE calculations, starting with pure $h$-BN. The cell structure was optimized with PBC, based on which the work function $\phi$ was computed. Parallelogram-shaped clusters with increasing sizes (Figure S1b) were also constructed, and  $I^\text{gas}$ was computed each cluster using the $\Delta$Kohn-Sham scheme.\cite{PhysRev.139.A619, TRIGUERO1999195} BE of the solid material ($I^\text{solid}$) was then computed using Eq. \ref{eq:cpm}.  Convergence in BE (Figure S1a) was reached at a size of 7$\times$7 supercell  (B$_{49}$N$_{49}$) with $a$=$b$=1.5 nm (Figure \ref{fig:Bc:fluo}a),  yielding a theoretical value of 189.84 eV. This value aligns well with the experiments (190.0--190.3 eV)\cite{guerra_2d_2018, achour_plasma_2017, zhou_high_2014, huang_exponentially_2021, deshmukh_ultrasound-assisted_2019, yu_thermal_2016, yang_thionyl_2021}. Further computational details and validations on cluster sizes are given in Supporting Information Notes 1 and 2. 

We proceeded to gradually\cite{radhakrishnan2017fluorinated} introduce 1-6  F atoms ($n_\text{F}=1, 2, \cdots, 6$) onto the 7$\times$7 supercell of pure $h$-BN, resulting in six F-BN structures (i-vi, Figure \ref{fig:Bc:fluo}b) with doping concentrations from 1\% (B$_{49}$N$_{49}$F) to 6\% (B$_{49}$N$_{49}$F$_{6}$). F(B) and F(N) are used to respectively denote a fluorine atom linked to a B and N sites, with $n_\text{F(B)}$ and $n_\text{F(B)}$ being the numbers in each F-BN model, such that 
\begin{equation}
\label{eq:nF}
n_\text{F}=n_\text{F(B)} + n_\text{F(N)}.
\end{equation}
These numbers of each configuration are summarized in Table S1.
Each supercell structure was re-optimized, at which the work function $\phi$ was computed (Table S2). BEs of borons at selected sites of i-vi were computed following the same settings as $h$-BN. The first F atom was always introduced to bond with boron in the center of the parallelogram model [denoted as B$_\text{c}$, see Figure \ref{fig:Bc:fluo}b (i)]. 

Figure \ref{fig:Bc:fluo}c shows the binding energies at B$_\text{c}$ during the gradual fluorination (B$\rightarrow$N$\rightarrow$B$\rightarrow$B $\rightarrow$N$\rightarrow$N)\cite{radhakrishnan2017fluorinated}, where sensitive influence was observed. Interestingly, fluorination at a  B (N) site consistently results in an additional decrease (increase)  of 0.14--1.06 (1.07--2.14) eV in the  BE of B$_\text{c}$, indicating evident competition. Comparison of the absolute values of the blue and red shifts indicates a stronger influence of  F(N) than F(B). Notably, when there are equal numbers of N- and B-site fluorination, configurations ii ($n_\text{F(B)}$=$n_\text{F(B)}$=1) and vi ($n_\text{F(B)}$=$n_\text{F(B)}$=3) consistently lead to a blue shift (1.07 and 3.08 eV, respectively) compared to $h$-BN. All calculations here show a range of 189.25-192.92 eV. One might argue that the numbers do not precisely coincide with the two representative experiments at 188.68\cite{venkateswarlu2019effective} and 191.0 eV\cite{radhakrishnan2017fluorinated}. Firstly, the experiment by Radhakrishnan \cite{radhakrishnan2017fluorinated}, as depicted in our Figure \ref{fig:exp}, pertains to low F concentrations. Measurement at much higher F concentrations\cite{radhakrishnan2017fluorinated} revealed a peak at approximately 195 eV (not included in our Figure \ref{fig:exp}). Secondly, our analysis here is only for one atom (B$_\text{c}$), whereas experiments detect statistical results (over many nonequivalent borons). Nevertheless, even at this stage of only B$_\text{c}$  calculations,  we have already clearly distinguished the contribution of different sites, and demonstrated that BE can be modulated through competitive fluorination at various sites.

\begin{figure}[!htb]
 \includegraphics[width=10cm]{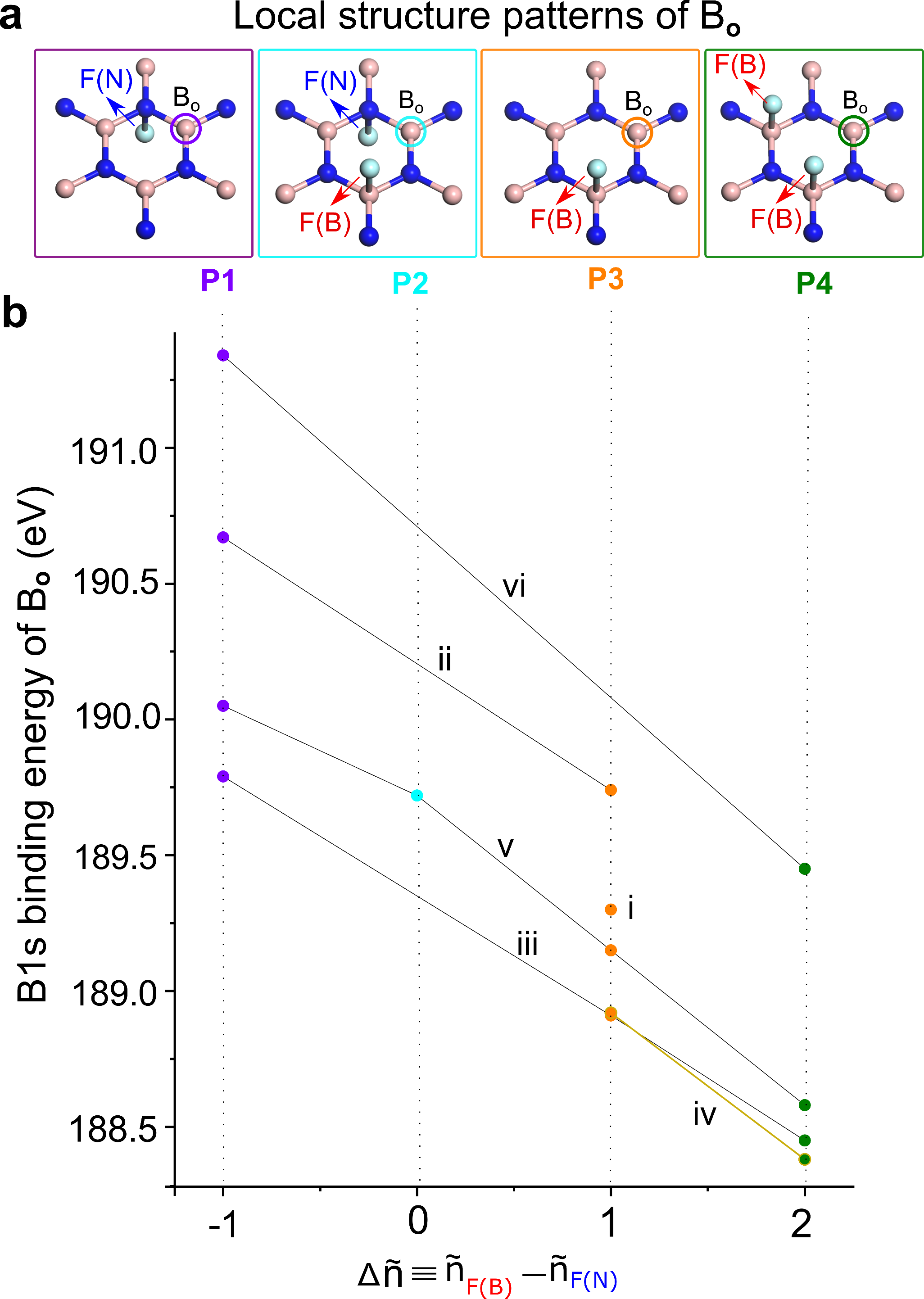}  
 \caption{
 (a) The local structures of selected B$_\text{o}$ atoms (unfluorinated borons; Figure S2) fall into four patterns P1-P4. F(N) and F(B) atoms in the nearest-neighborhood are labeled. Their number difference ($\tilde{n}_\text{F(B)}- \tilde{n}_\text{F(N)}$) is defined as $\Delta\tilde{n}$ (all numbers are provided in Table S3). (b) Relation between simulated B1s BE of B$_\text{o}$ in F-BN configurations i-vi and $\Delta \tilde{n}$.
  }
\label{fig:Bo:nonF}
\end{figure}

Then, more borons were investigated to see the BE change induced by fluorination. Unfluorinated borons (B$_\text{o}$; Figure S2) near the fluorinated center were selected for BE calculations, and results are presented in Figure \ref{fig:Bo:nonF},  ranging from 188.38 to 191.34 eV. The lower and upper ranges closely align with experimental values of 188.68 \cite{venkateswarlu2019effective} and 191.0 eV\cite{radhakrishnan2017fluorinated}, respectively. In a real 2D F-BN nanosheet, there are always more B$_\text{o}$ than B$_\text{c}$ atoms at low fluorination concentrations.

We categorized the nearest-neighboring unfluorinated borons, B$_\text{o}$, in all six F-BN models into four patterns (P1-P4; Figure \ref{fig:Bo:nonF}a). This classification was guided by whether the nearest-neighboring borons and nitrogens of B$_\text{o}$ were fluorinated or not. Within each pattern, we use $\tilde{n}_\text{F(B)}$ [$\tilde{n}_\text{F(N)}$]  to represent the total number of fluorine atoms attached to the nearest-neighboring boron (nitrogen) of B$_\text{o}$ (cf. $n_\text{F(B)}$ and $n_\text{F(N)}$, meaning numbers in each F-BN configuration), and the difference is given by,
\begin{equation}
\label{eq:deltan}
\Delta \tilde{n} \equiv \tilde{n}_\text{F(B)} - \tilde{n}_\text{F(N)}.
\end{equation}
These numbers in P1-P4 are summarized in Table S3.  In Figure \ref{fig:Bo:nonF}b, we observed correlations of $\Delta \tilde{n}$ and the B1s BEs of B$_\text{o}$. We found that within the same F-BN, boron with a larger $\Delta \tilde{n}$ corresponds to a lower BE. This suggests that fluorination has similar effects on BEs of both B$_\text{c}$ and B$_\text{o}$.  BE of B$_\text{o}$ is also the result of competitive interactions between F(B) and F(N), which have opposite effects.

\begin{figure*}[!htb]
\centering
 \includegraphics[width=15cm]{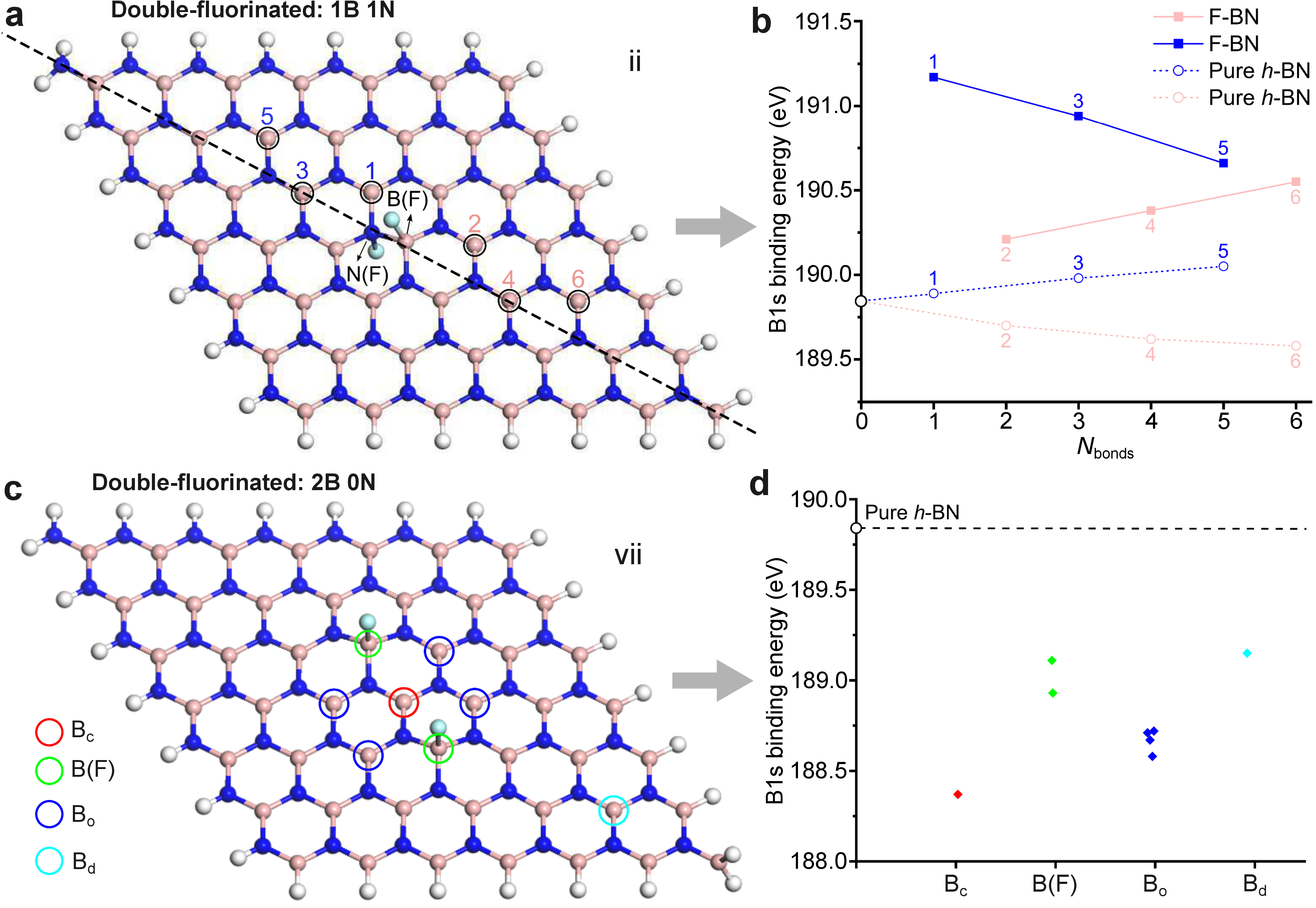}  
 \caption{Theoretical B1s BEs of two double-fluorinated F-BN configurations: (a-b)  ii, (c-d)  vii.
(a) Structure of ii, where 1--3 (4--6) label unfluorinated borons in the diagonal (B$_\text{d}$). They refer to number of bonds, $N_\text{bonds}$, away from the fluorinated N (fluorinated B). (b) Simulated B1s BEs of B$_\text{d}$  with respect to $N_\text{bonds}$. (c) Structure of vii. B$_\text{c}$, B(F), B$_\text{o}$, and B$_\text{d}$, are labeled in circles. (d) Simulated B1s BEs for borons in panel c. Dashed line refers to the simulated BE of pure $h$-BN (189.84 eV).
 }
\label{fig:2F}
\end{figure*}

To explore the influence range of fluorination, we calculated BEs of six borons (Figure \ref{fig:2F}a) along the diagonal line (B$_\text{d}$) of F-BN structure ii. This configuration encompasses scenarios with fluorinated centers at a boron [B(F)] and a nitrogen [N(F)] situated near the center.  Among the six borons, three are positioned on the B(F) side with distances of 2, 4, and 6 chemical bonds, respectively, while the remaining three are situated on the N(F) side with distances of 1, 3, and 5 bonds. It is important to consider the potential impact of edge effects imposed by the size limit of our structural model. To address this concern and accommodate any deviation it may introduce, we also calculated BEs of B$_\text{d}$ at the corresponding positions within pure $h$-BN as a ``baseline'' for comparison.

Figure \ref{fig:2F}b illustrates simulated BEs of B$_\text{d}$ in structure ii and pure $h$-BN.   In structure ii, BEs of B$_\text{d}$ on the B(F) [N(F)] side progressively increases by 0.34 eV (decreases by 0.51 eV) along the diagonal line away from each fluorination center.    The monotonic increment and decrement indicate that F(B) and F(N) primarily impact borons on their respective sides.    As the distance from B(F) and N(F) increases, the influence of F(B) and F(N) on BEs of B$_\text{d}$ gradually diminishes.     Consequently, BEs of B$_\text{d}$ on the B(F) side gradually recovers, while BEs on the N(F) side declines, eventually converging at ca. 190.60 eV.     This indicates that when B$_\text{d}$ is sufficiently distant from the fluorinated center, the effects of F(B) and F(N) on its BE become relatively indistinguishable.      However, it should be noted that BEs on both sides do not converge to pure $h$-BN's BE of 189.84 eV, but were about 0.76 eV higher than this value, indicating a global impact of fluorination on all atoms within the system. A comparable effect of fluorination on the C1s BEs of all carbons in molecules was reported very recently. \cite{bischof_chemical_2023} Meanwhile, in pure $h$-BN, BEs of B$_\text{d}$ on the B(F) [N(F)] side decrease by 0.12 eV (increase by 0.16 eV) as they move along the diagonal line away from the fluorinated center, estimating the edge effect.  Interestingly, the evolution in $h$-BN and F-BN always occurs in opposite directions, indicating that the observed conclusion is not a side effect of limited model size but is a result of fluorination.

Furthermore, we computed a new configuration vii fluorinated at two borons (Figure \ref {fig:2F}c).  Seven borons were selected covering the four types, including 1 B$_\text{c}$, 2 B(F), 4 B$_\text{o}$, and 1 B$_\text{d}$ atoms.   Computed BEs are presented in Figure \ref {fig:2F}d.  We found that B1s BEs of the two double-fluorinated  structures ii and vi are significantly different. BEs of all selected borons in vii (188.37-189.15 eV) are much lower than that of pure $h$-BN, aligning well with Venkateswarlu et al.\cite{venkateswarlu2019effective} This decrease in BE is directly attributed to the presence of F(B). Even though B$_\text{d}$  is far away from the fluorination center, its BE is still 0.69 eV lower than pure $h$-BN, again indicating that fluorination impacts the entire F-BN system. This opposes  the transferability from $h$-BN to F-BN, and it is not appropriate to simply assign a deconvoluted peak at 190.0-190.3 eV as originating from unfluorinated borons.\cite{radhakrishnan2017fluorinated}

In summary, we conducted a comprehensive DFT study of B1s XPS of seven F-BN nanosheets (i-vii), doped with 1-6 F atoms (equivalent to 1-6\% in concentration) to a 7$\times$7 supercell of pure $h$-BN (B$_\text{49}$N$_\text{49}$). The aim was to understand the impact of fluorination on the 1s BEs of different boron sites.   Our calculations revealed that F(B) and F(N) exert a competitive effect on the 1s binding energies of borons.  We further elucidated that the peak observed at 188.68 eV (a red shift compared to $h$-BN) by Venkateswarlu et al.\cite{venkateswarlu2019effective} is due to the predominant binding of F atoms to B atoms, while the peak reported\cite{radhakrishnan2017fluorinated, ahmad_experimental_2020, ul_ahmad_novel_2019} at 190.37 eV (indicating a slight blue shift), occurs when F atoms bind non-exclusively to both B and N atoms.     Our calculations have confirmed a significant impact of fluorination on all atoms in the $\pi$-conjugated material, indicating poor transferability of B1s BE between $h$-BN and F-BN. Consequently,  attributing a Gaussian deconvoluted peak at 190.0-190.3 eV in pure $h$-BN to signify an unfluorinated boron contribution is not appropriate. To clarify the structure-XPS relation and facilitate controllable fluorination, direct and precise simulations are essential.


\section*{Supporting Information}
The Supporting Information is available free of charge at https://pubs.acs.org/doi/XXXX/XXXXXX.\\
Supplementary computational  details; 
Validation of spectral convergence in pure $h$-BN; 
Figure for selected B$_o$ atoms in i-vi; 
Figure for enlarged view of ii and vii;
Tables for numbers of F atoms in i-vi and P1-P4;
Table for computed work functions and bond lengths; 
Cartesian coordinates of all optimized structures (PDF).

\section*{Author Contributions}
W.H. led the project and proposed the idea. 
Y.X. performed all calculations with the help of J.-R.Z and S.-Y. W.
J.-R.Z assisted the data analysis and contributed to the results explanation. 
Y.X., J.-R.Z, and W.H wrote the manuscript. 
All authors discussed the results and commented on the manuscript.

\section*{Notes}
The authors declare no competing financial interest.\\
All computational details and data supporting the findings of this study are available within the main text and its Supporting Information.
The data are also available from the corresponding authors upon reasonable request.

\section*{Acknowledgements}
Financial support from the National Natural Science Foundation of China (Grant No. 12274229) is greatly acknowledged.

\providecommand{\latin}[1]{#1}
\makeatletter
\providecommand{\doi}
  {\begingroup\let\do\@makeother\dospecials
  \catcode`\{=1 \catcode`\}=2 \doi@aux}
\providecommand{\doi@aux}[1]{\endgroup\texttt{#1}}
\makeatother
\providecommand*\mcitethebibliography{\thebibliography}
\csname @ifundefined\endcsname{endmcitethebibliography}
  {\let\endmcitethebibliography\endthebibliography}{}

\end{document}